\documentclass[aps,pra,twocolumn,a4paper,superscriptaddress,amsmath,amssymb,floats,floatfix,english]{revtex4}
\usepackage{graphicx}
\usepackage{amsmath}
\usepackage{amssymb}
\usepackage{braket}

\usepackage[T1]{fontenc}
\usepackage{ifthen}
\usepackage{times}
\usepackage{babel}
\usepackage{bm}
\usepackage{pstricks}
\usepackage{times}
\usepackage{soul}
\usepackage{color}
\usepackage{placeins}

\usepackage{hyperref}       
\usepackage{url}            
\usepackage{booktabs}       
\usepackage{amsfonts}       
\usepackage{nicefrac}       
\usepackage{microtype}      
\usepackage{lipsum}
\usepackage{xcolor}
\usepackage{wrapfig}
\usepackage{floatflt}
\usepackage{float}

\newcommand{\REM}[1]{\ifthenelse{0=1}{#1}{}}

\begin{document}
\title{Quantum versus classical angular momentum }

\author{Jan Mostowski}
\affiliation{Institute of Physics of the Polish Academy of Sciences, Al. Lotnik{\'o}w 32/46, 02-668 Warszawa, Poland
}
\email{mosto@ifpan.edu.pl, pietras@ifpan.edu.pl}
\author{Joanna Pietraszewicz}
\affiliation{Institute of Physics of the Polish Academy of Sciences, Al. Lotnik{\'o}w 32/46, 02-668 Warszawa, Poland
}
\date{\today}
\begin{abstract}
Angular momentum in classical mechanics is given by a vector. The plane perpendicular to this vector, in accordance to central field theory, determines the space in which particle motion takes place. No such  simple picture exists in quantum
mechanics. States of a particle in a central field are proportional to spherical harmonics which do
not define any plane of motion. In the first part of the paper we discuss the angular distribution of
particle position and compare it to the classical probabilistic approach. In the second part, the matter of addition of angular momenta is discussed. In classical mechanics this means addition of vectors while in quantum mechanics Clebsch–Gordan coefficients have to be used. We have found classical approximations to quantum coefficients and the limit of their applicability. This analysis gives a basis for the so called "vector addition model" used in some elementary textbooks on atomic physics. It can help to understand better the addition of angular momenta in quantum mechanics.  
\end{abstract}


\maketitle

\section{Introduction}
Angular momentum in classical mechanics measures the "amount of rotation". In a sense, it is analogous to linear momentum, which measures the "amount of motion". 
The exact definition of angular momentum is usually given in undergraduate physics courses. For a point particle, it is defined
as the cross product of position and momentum vectors. A standard reasoning leads to the conservation law, according to which angular momentum is conserved in case of motion in a central field. The particle motion becomes then restricted to a plane perpendicular to the angular momentum vector.

 Turning now to quantum mechanics, one needs operators representing physical quantities and states (wavefunctions) specifying the system. Angular momentum is the physical quantity considered in this paper.
 Operators representing this quantity are defined  as the cross product of position and momentum operators. 
 Like in classical mechanics, the angular momentum is conserved in case of central interactions. Unlike in classical mechanics, however, the plane of motion is not uniquely specified.

Note also, that no state can be a common eigenstate of all three vector components of angular momentum. 
Common eigenstates of the total angular momentum and of one of its components, usually the $z$ component, exist and are routinely  used in the description of systems with spherical symmetry. 

States with maximal and minimal values of the $z$ component of angular momentum, i.e. equal to the total angular momentum $j$ or $-j$, correspond to motion in the $xy$ plane. This property is stated in some textbooks \cite{Shankar}, and educational papers \cite{Pitak}. States with other values of $z$ component are much more difficult to interpret. 

Inspection of eigenfunctions of the total angular momentum and the $z$ component (spherical harmonics) and their
dependencies on angles in spherical coordinates, can be confusing. It is very difficult to find resemblance between these states and more familiar quantities known from classical mechanics, e.g. the plane of motion. Thus the quantum -- classical correspondence angular momentum states remains unclear to many students. 

Addition of angular momentum in quantum mechanics is even more confusing. Addition of angular momenta in classical mechanics is very simple -- two vectors of angular momenta should be added to get the total angular momentum. The corresponding case in quantum mechanics is not that simple. Standard textbook explanation of addition of angular momentum does not resemble vector addition. Addition of angular momenta requires introduction of Clebsch-Gordan coefficients. Explicit formulas for them are usually restricted to small values of angular momentum and their relation to the addition of classical angular momenta is usually not discussed in textbooks. 

In the present paper we formulate classical probabilistic models that mimic quantum states with a given angular momentum. These models originate form the so called "vector model" of angular momentum, used in some textbooks on atomic physics \cite{Friedman11,Haken, Saari16}. As opposed to these textbooks, we provide a thorough discussion of these models, compare them with full quantum mechanical treatment and discuss the range of applicability. In this way the relation between the classical and quantum approach to the angular momentum is established and visualized. 
Classical models are also supported by semiclassical (WKB) approximations to the angular momentum states. 

Further, we use a classical probabilistic approach to angular momenta addition to compare it with the quantum case, and then discuss the range of applicability of classical approach. All this helps students to understand the relation between classical vector addition of angular momenta and the corresponding quantum case. 

We should also mention that quantum systems with large angular momenta are a subject of studies in recent years, for~ e.g.~ \cite{Boyd14, Zeilinger16, ournew}. Deep understanding of these states and their relation to classical physics seems to be an important part of advanced physics education. The probabilistic approach presented here can be used in courses on quantum mechanics at the graduate level. The results presented should help students to better understand this element of quantum mechanics.

\section{Classical approach to angular momentum}
We will begin with formulation of purely classical motion and classical angular momentum. 
Consider a point particle with mass~ $\mu$ undergoing circular motion in a plane. Orientation of the  plane is characterized by the angular momentum vector, which forms angle\, $\beta$ with the\, $z$ axis. 

It is of interest to describe the particle motion by giving the time dependence of the its coordinates. In the plane of motion, i.e. perpendicular to the angular momentum, the~ motion is given by\  $x'(t)=r\cos(\omega t+\alpha)$, $y'(t)=r\sin(\omega t+\alpha)$, where\,  $x'$ and\,  $y'$ are in-plane coordinates, and \, $r$ is the radius of the orbit. Angle $\alpha$ denotes the initial phase of the motion. 

\begin{figure}[ht]
    \includegraphics[width=6cm]{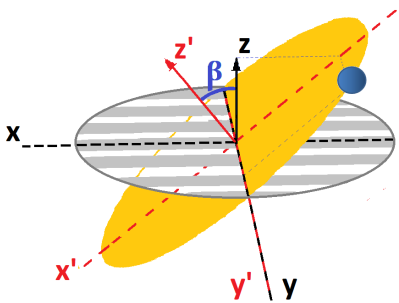}
    \caption{Trajectory of a particle lies in the $x',y'$ plane. Angular momentum vector is perpendicular to the plane of motion, hence is directed along the $z'$ axis, which forms angle $\beta$ with the $z$ axis of the~ laboratory coordinate frame. The $y$ axis in the laboratory frame and the $y'$ axis in the plane of motion are common for both frames. The dotted blue lines show projection of particle position on $x'$ ad $y'$ axes.
    }
    \label{fig:fig1}
\end{figure}

In order to find particle coordinates  in the laboratory frame one has to rotate the coordinate system around\ $y$ axis by angle~ $\beta$. After such rotation one gets:
\begin{eqnarray}
 x(t) &=& x'(t)\, \cos{\beta}= r\, \cos(\omega t+\alpha)\, \cos{\beta} \\
 y(t) &=& y'(t) = r\, \sin(\omega t+\alpha) \\
 z(t) &=& x'(t)\, \sin{\beta}= r\, \cos(\omega t +\alpha)\, \sin{\beta}.
\label{motion z}
\end{eqnarray}
The system together with coordinates is presented in Fig. \ref{fig:fig1}. 

The value of the angular momentum $L$ is related to the~ parameters $r$ and $\omega$ by formula\  $L=\mu r^2\omega$, and the $z$ component of angular momentum is  $L_z=\mu r^2\omega^2\cos{\beta}$.

We have to use probabilistic approach to this classical theory,  since this is required by quantum mechanics to be able to compare both descriptions in the next sections. Therefore, we introduce ensemble of particles to classical theory. Motion of each particle is characterized by \ $\alpha$\, which is a
random variable with uniform distribution in the range\  $[0 ,\, \pi]$. We will look for the probability density that the position vector forms angle~ $\theta$ with the\ $z$ axis, i.e.,  
\begin{equation}
    p(\cos\theta)=\frac{1}{2\pi}\int_0^{2\pi} d\alpha\ \delta\left(\cos\theta - \frac{z(t)}{r}\right).
\end{equation}
Inserting value of\ $z(t)$ from\,  Eq.(\ref{motion z}) and integrating over $\alpha$, we~ get:
\begin{eqnarray}
    & p(\cos\theta)=
    \frac{1}{2\pi}\frac{1}{\sqrt{\cos^2{\theta}-\sin^2{\beta}}} \nonumber \\
   &   \mbox{in the "allowed range" }\quad  |\cos\theta| > |\sin\beta|,
    \label{theta}
\end{eqnarray}
or\quad $p(\cos{\theta})=0$\quad for\ $\cos{\theta}$\quad outside the allowed range. 
Further discussion of the function $p(\cos{\theta})$ and comparison with its quantum analogue will be given in the next section.

\section{Angular momenta in quantum mechanics}
Description of any physical system in the framework of~ quantum mechanics requires information about the state of the system, and about relevant physical quantities represented by linear operators. In this paper the relevant physical quantities are components of angular momenta, we will not consider any other physical quantities. 

Angular momentum operators\  $J_x$, $J_y$, and\ $J_z$ obey commutation relations:
\begin{equation}
    [J_{\sigma},J_\nu]=i\hbar\ \epsilon_{\sigma \nu\kappa}\ J_\kappa\ ,
    \label{commutations}
\end{equation}
where\ $\epsilon_{\sigma \nu\kappa} $\  is the purely antisymmetric unit tensor. The states of the system can be chosen as common eigenstates of the operator\ $J^2=J_x^2+J_y^2+J_z^2$ \big(with eigenvalues\  $\hbar^2 j(j+1)$, where $j=0,1,\dots$\big)\, and operator \ $J_z$ \big(with eigenvalues $\hbar m$, where $m=-j,-j+1,\dots j$\big). These states are denoted~ by~ $|j,m\rangle$. 

Observe that two other components of the angular momentum, $J_x$ and $J_y$, do not have well defined values in states\ $|j,m\rangle$. Their average values are equal to zero, whereas their dispersions are equal to
\begin{equation}
    \Delta^2 J_x = \Delta^2 J_y =\frac{\hbar^2}{2}\,  \Big( j(j+1)-m^2\Big)\ .
\end{equation}
States with\ $m=\pm j$\  have the smallest dispersion. States with\ $m=0$\  have the largest dispersion, comparable to\ $j$.

Angular momentum operators are often considered in the position representation: 
\begin{equation}
    J_\sigma=\frac{\hbar}{i}\ \epsilon_{\sigma\nu\kappa}\  x_\nu \ \frac{\partial}{\partial x_\kappa}\ ,
\end{equation} 
where\ $x_{\kappa}$, with\  $\kappa = 1,2,3$, denote cartesian coordinates. In this representation the eigenstates of operators\ $J^2$\ and\ $J_z$\ are given by spherical harmonics $Y_{j,m}(\theta,\varphi)$. There are many different phase conventions for $Y_{j,m}(\theta,\varphi)$. We have adopted the phase convention used in  Mathematica. 

Explicit form of spherical harmonics can be found in many textbooks, so we will not reproduce them here. We will, however, mention a peculiar feature possessed by  spherical harmonics $Y_{j,j}(\theta,\varphi)$. Since they are proportional to~\ $\sin^j(\theta)\ \exp(i\, j\, \varphi)$,\  they strongly concentrate around $\theta \approx 0$ in the limit of large $j$. This is an argument for treating 
 $Y_{j,j}(\theta,\varphi)$ as state which is an analogue of classical motion in~ the~ $\theta=0$, hence in $xy$, plane \cite{Shankar, Pitak}. Other spherical harmonics do not have this property. They are not concentrated in a plane and, moreover, they exhibit oscillations with\ $\theta$.  One~ of~ the~ aims of the paper is to provide classical interpretation for these states.
 
\begin{figure}[ht]
    \centering
    \includegraphics[width=6cm]{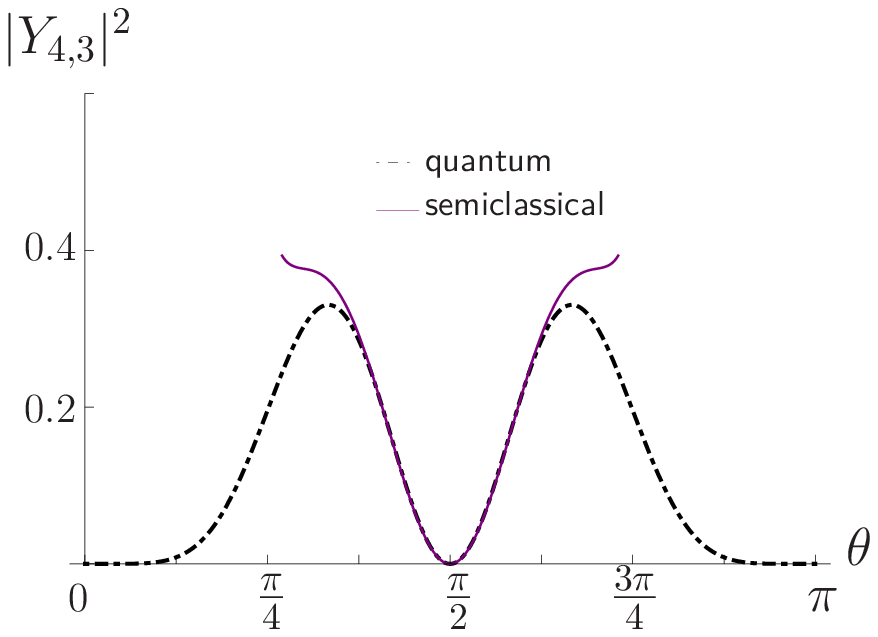}
    \includegraphics[width=6cm]{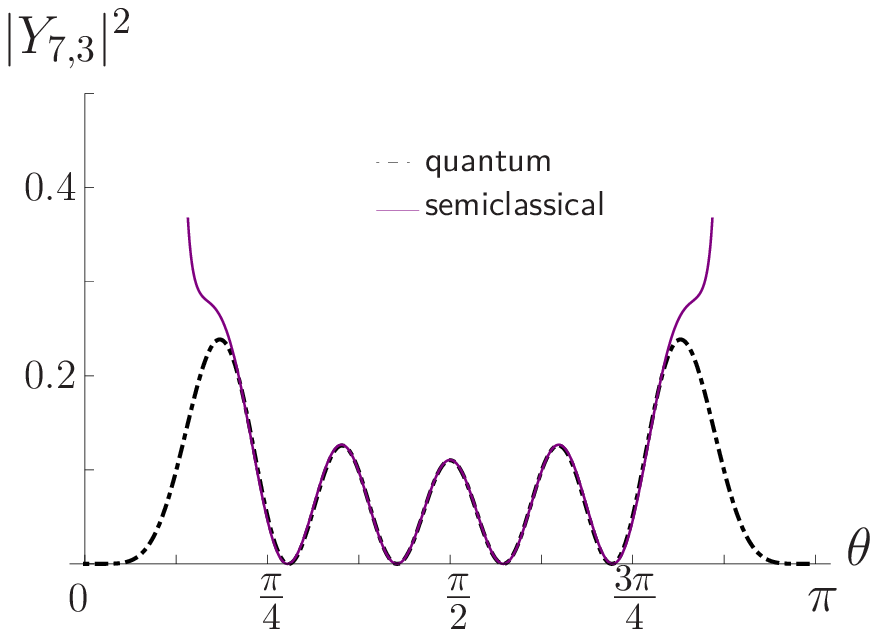}\\
    \includegraphics[width=6cm]{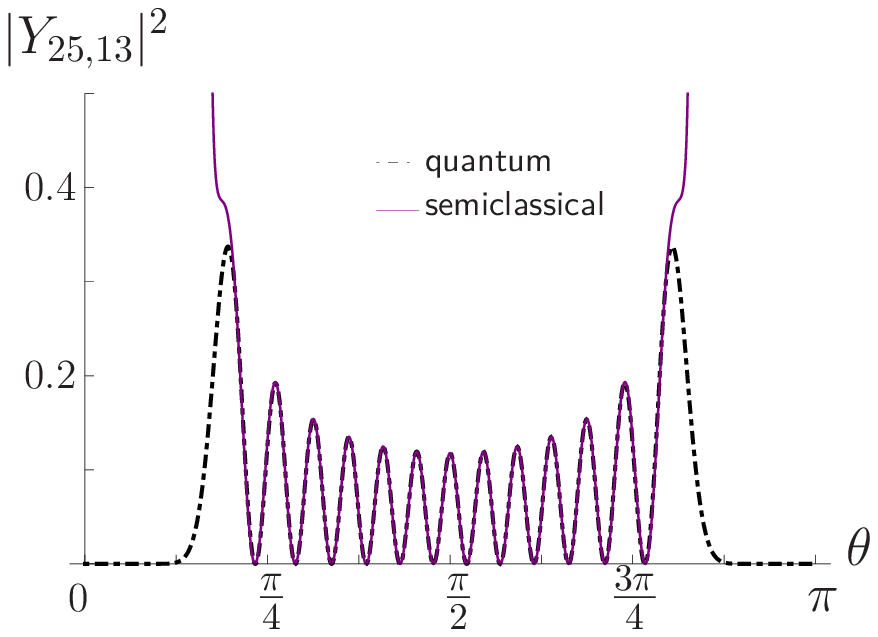}
    \includegraphics[width=6cm]{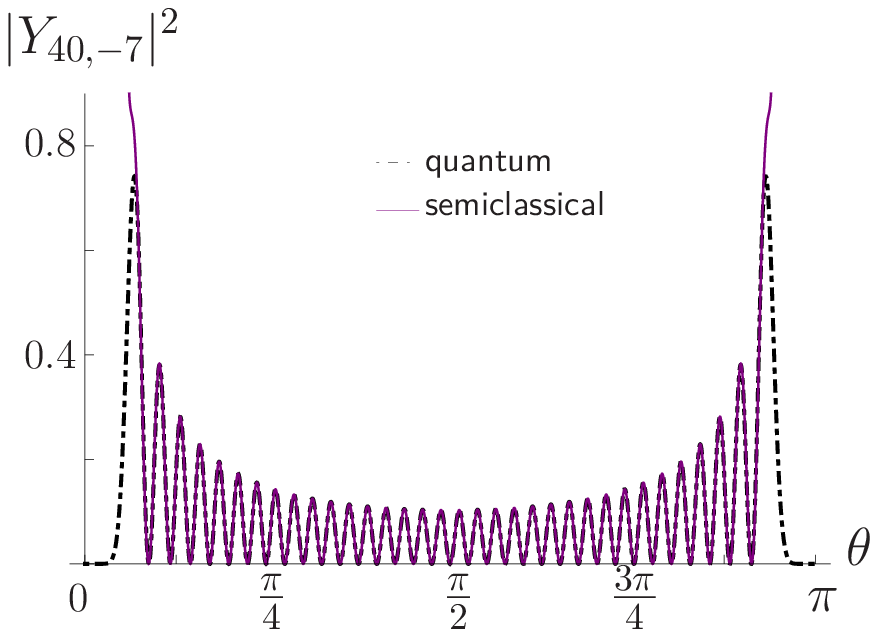}
    \caption{Spherical harmonic squared $|Y_{j,m}(\theta,\varphi)|^2$ versus angle $\theta$ (the modulus square does not depend on $\varphi$) are presented for quantum solution (black, dot-dashed line) and for its semiclassical counterpart (purple, solid line). Values $j$ and $m$ are on the plot. Classical angular momentum in all figures are given in units of $\hbar$.}
    \label{fig:spherical}
\end{figure}

 Absolute values squared of some spherical harmonics are shown in \ Fig.~\ref{fig:spherical}. In addition, their semiclassical (WKB) approximation valid for large $j$ are also given. For details of the semiclassical approximation see  Supplementary material (\ref{supwkb}).  
 
 Semiclassical approximation is introduced here to give the feeling of the overall behavior of spherical harmonics. If one interprets $j$ classically as the total angular momentum, and $m$ as its $z$ component, then the classical motion is restricted to the interval\ $-\frac{|m|}{j}\, \le \, \sin{\theta}\,  \le\,  \frac{|m|}{j}$. Thus "classical turning points" should exist also in the quantum case. One can clearly see these points in case of large $j$ and $m$ -- spherical harmonics have large values around such $\theta$ that $m^2\approx j^2 \sin^2\theta$. Beyond "classical turning points"  spherical harmonics decay to zero, while in the "classically allowed region" they exhibit oscillations. The period of these oscillations scales with $j$ as $j^{-1}$.  

One can interpret the 
modulus square of the spherical harmonics $|Y_{j,m}(\theta,\varphi)|^2$ as the probability density of finding a particle with quantum numbers $j$ and $m$ at the angle $\theta$. This probability density can be compared to purely classical result, namely  Eq.(\ref{theta}), which gives the classical probability distribution of angles. Firstly, we see that the classical distribution mimics the quantum one in case of large $j$. The exception are rapid oscillations, that are present in the quantum case and absent in the classical distribution. Secondly, we see that the classical approximation gives the average over several oscillations of the quantum result. In other words, classical approach is valid if the resolution of the measuring device does not allow to measure high frequency oscillations.

One should bear in mind, however, that the classical angular momentum is a continuous variable as opposed to the quantum case. We have identified the classical value of $L$ with the quantum value $\hbar j$, and the classical $L_z$ with the magnetic quantum number $\hbar m$, although whole ranges of  $L$ and of $L_z$, with lengths comparable to $\hbar$, approximate equally well the quantum case. 

\section{Addition of angular momenta}
We will  begin by discussing addition of
angular momenta in classical mechanics. Consider two spinning tops, one with angular momentum\  ${\bf L}_1$, the other with angular momentum~\ ${\bf L}_2$. The total angular momentum of the system is thus\  ${\bf L}~=~{\bf l}_1+{\bf l}_2 $. The length of the total angular momentum is\quad $L~=~\sqrt{L_1^2+L_2^2+2\, L_1L_2\, \cos{\alpha}}$, where\ $L_1$ and\ $L_2$\ denote lengths of the appropriate vectors and\ $\alpha$ is the angle between them. The range of \ $L$\ values extends between\ $\left|L_1-L_2\right|$\  and\  $(L_1+L_2)$, depending on the angle $\alpha$ between the two vectors. The direction of the total angular momentum can be easily found with the help of vector addition.

We will turn now to quantum mechanics. Consider two subsystems, one in the state $|j_1,m_1\rangle$ and the second one in~ the~ state $|j_2,m_2\rangle$. These states belong to different spaces,   
so the space of states of the combined system is the (tensor) product of two spaces.
Among all states of the whole system one can distinguish product states
$|j_1,m_1\rangle |j_2,m_2\rangle$.
The meaning of this states is that the first subsystem is in the $|j_1,m_1\rangle$ state and the second subsystem  in the $|j_2,m_2\rangle$ state. 

The total angular momentum operator is defined, as in classical physics, by the sum of individual components, $J_\sigma~=~J_{1\sigma}~+~J_{2\sigma}$, where $\sigma=x,y,z$. 
The operators\ $J_{1\sigma}$, $J_{2\sigma}$\  satisfy the same 
the commutation relations as the individual components, see Eq.(\ref{commutations}).

The product states are not, in most cases, eigenstates of the square of the total angular momentum ${\bf{J}}^2=J_x^2+J_y^2+J_z^2$,  
but definitely they are eigenstates of $J_{1z}+J_{2z}$. 
One can find, however, states in the product space which 
are eignestates of both ${\bf J}^2$ and $J_z$. They are denoted by\ $|J,M\rangle$\ and are, of course, linear combinations of the product states:
\begin{equation}
    |J,M\rangle=\sum_{m_1,m_2}\ \textrm{CB}(j_1,m_1,j_2,m_2;J,M)\quad |j_1,m_1\rangle\ |j_2,m_2\rangle.
\end{equation}
Coefficients CB$(j_1,m_1,j_2,m_2;J,M)]=\textrm{CB}$ are called Clebsch-Gordan coefficients. Explicit formulas for them can be found in~ some textbooks, but these formulas are not very useful and will not be reproduced here.
Values of CB$(j_1,m_1,j_2,m_2;J,M)$, when needed, can be found in  quantum mechanics textbooks or 
in easily available tables. Some symbolic computer languages, like Mathematica, have build-in procedures to find their values. 

We have to bear in mind that Clebsch-Gordan coefficients are probability amplitudes and that there is no way to define their phases in an unambiguous way. Most, if not all, textbooks use the so called Shockley convention where all coefficients are real, while their signs are fixed explicit.

Explanation of the overall behavior of the CB coefficients based on a classical model will be given in the next section.

\section{Classical and quantum Clebsch-Gordan coefficients}
\begin{figure}[ht]
    \includegraphics[width=6.3cm]{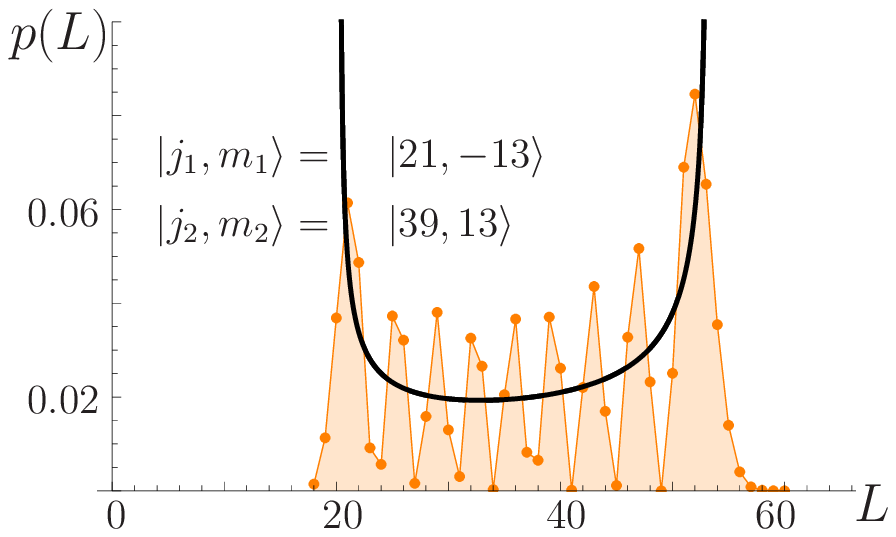}
    \includegraphics[width=6.3cm]{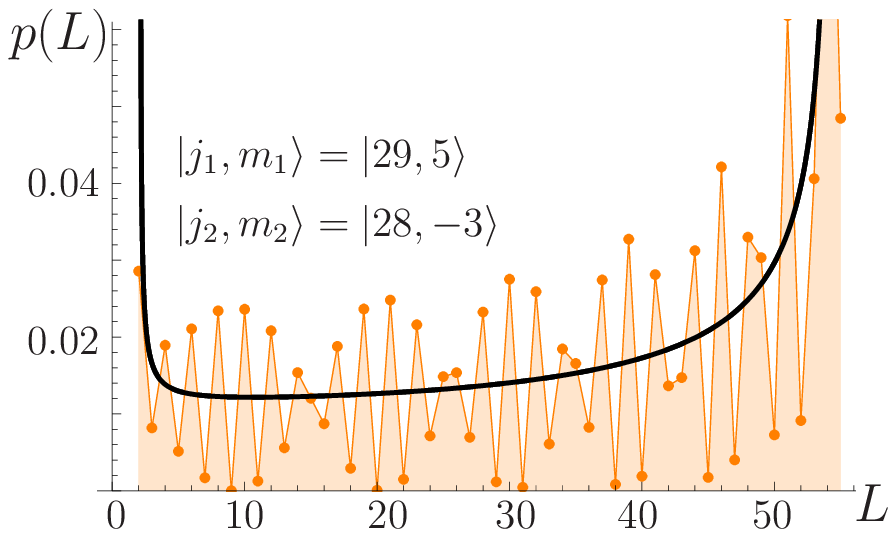}
    \includegraphics[width=6.3cm]{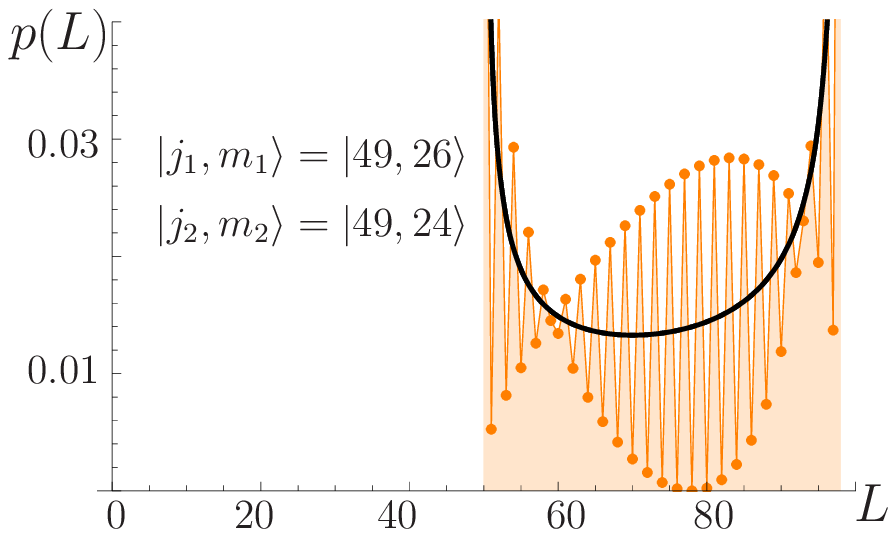}
    \includegraphics[width=6.3cm]{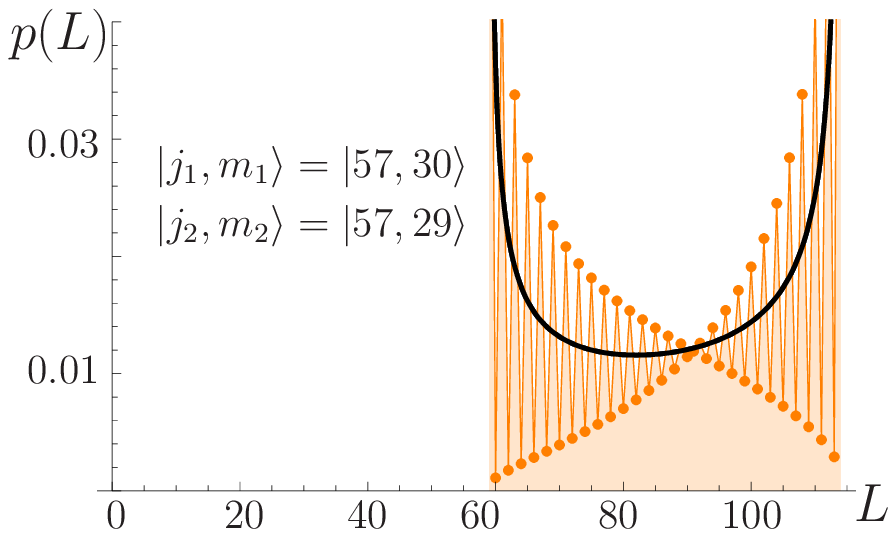}
    \caption{Probability distribution of the total angular momentum (equivalent to  the probability of the azimuth angle). The orange points denote values of the squares of Clebsch-Gordan coefficients that fulfill conditions imposed by the quantum numbers $j_1,m_1,j_2,m_2$. Their values are shown in each plot.
    The solid lines are semi-classical results. The orange area  has no physical meaning.} 
    \label{fig:fig3}
\end{figure}

We will formulate a classical probabilistic model of addition of two angular momenta and compare the results with the exact quantum values. 

The classical model is supposed to mimic addition of quantum angular momenta. The quantum\ $|j,m\rangle$\ states, that will be superimposed and their angular momenta added, have a~ well defined total angular momentum and the $z$ component of angular momentum, while the other two components are random -- loosely speaking. This gives an inspiration for using the classical approach, formulate below. We follow the ideas from the Wigner book \cite{Wigner}.

Let us assume that the classical angular momentum  vector~\  ${\bf L}_1$\ has components
\begin{eqnarray}
L_{1,x}&=&L_1\sin{\theta_1}\cos{\varphi_1}, \nonumber\\ 
L_{1,y}&=&L_1\, \sin{\theta_1}\sin{\varphi_1},\nonumber \\
L_{1,z}&=&L_1\, \cos{\theta_1}.\nonumber
\end{eqnarray}
The components of the second vector\  ${\bf L}_2$\ are
\begin{eqnarray}
L_{2,x}&=&L_2\, \sin{\theta_2}\cos{\varphi_2}, \nonumber\\
L_{2,y}&=&L_2\, \sin{\theta_2}\sin{\varphi_2}, \nonumber\\
L_{2,z}&=&L_2\, \cos{\theta_2}. \nonumber
\end{eqnarray}
Angles\ $\theta_1$\ and\ $\theta_2$\ define the $z$ components of both angular momenta and are assumed to be fixed. Angles\ $\varphi_1$\ and\ $\varphi_2$, defining the $x$ and $y$ components of angular momenta, are assumed to be random variables with uniform distribution.

We will now find the probability distribution of the square of the total angular momentum. It is given by the average value of the\ $\delta$ function over possible angle settings:
\begin{equation}
p(L^2)=\frac{1}{(2\pi)^2}\int \ d\varphi_1 d\varphi_2\
\delta\Big(L^2-L_1^2-L_2^2-2\, L_1 L_2\cos{\alpha}\Big). 
\label{distribution}
\end{equation}
The angle\ $\alpha$ between the vectors can be expressed in terms of $\theta_1$, $\theta_2$, $\varphi_1$ and $\varphi_2$:
\begin{equation}
\cos{\alpha}=\cos{\theta_1}\cos{\theta_2}
+\, \sin{\theta_1}\sin{\theta_2}\,  
\cos\left(\varphi_1-\varphi_2\right).
\label{angles}
\end{equation}
\begin{widetext}
Inserting Eq.(\ref{angles}) into Eq.(\ref{distribution})  
we get
\begin{equation}
p(L^2)=\frac{1}{2\pi}
\int\, d\varphi\ \delta\Big(L^2-L_1^2-L_2^2-2 L_1 L_2\, \left(\cos\theta_1 \cos{\theta_2}
+\sin{\theta_1}\sin{\theta_2}\cos\varphi\right)\Big). \label{distribution2}
\end{equation}
The integration over\ $\varphi=\varphi_1-\varphi_2$\  gives
\begin{equation}
    p(L^2)= \frac{1}{\pi}\ \frac{1}{\sqrt{\Big(L_1^2+L_2^2+2L_1 L_2\, \cos{(\theta_1+\theta_2)-L^2}\Big)\Big(L^2-L_1^2-L_2^2-2L_1 L_2\, \cos{(\theta_1-\theta_2)}\Big)}}.
\end{equation}
\end{widetext}
Finally, the distribution of $L$ can be obtained  with the help of the relation
\begin{equation}
    p(L)=2 L\ p(L^2).
\end{equation}

To make the quantum-classical correspondence even more readable we have to interpret angles\  $\theta_1$ and\ $\theta_2$. Interpretation should be just like in quantum mechanics, i.e., the projection of the angular momentum on the $z$ axis is equal to the magnetic quantum number\  $m$. Therefore,\  $L_{z_{1}}=L_1\cos{\theta_1}$,\ $L_{z_{2}}=L_2\cos{\theta_2}$\ and\ $L_z=L_{z_{1}}+L_{z_{2}}$, which allows ones to write:
\begin{eqnarray}
    p(L)&=&\frac{1}{\pi}\frac{2L}{\sqrt{\mathcal{A}^2+4\ \Big[L^2\,  L_{z_1}\, L_{z_2}-\left(L_2^2\, L_{z_{1}} + L_1^2\, L_{z_2}\right)\, L_z \Big] }} \nonumber \\
      \\
    &&\mathcal{A}^2=-L^4 - L_1^4 - L_2^4 + 2 L^2 \Big(L_1^2 + L_2^2\Big)  + 2 L_1^2 L_2^2 \nonumber \\
    \label{classical_distribution}
\end{eqnarray}

The above probability distribution $p(L)$ should be understood as the classical equivalent of the CG coefficient squared. We have got, therefore, a classical analogue of the Clebsch-Gordan coefficients.

\begin{figure}[ht]
    \centering
    \includegraphics[width=7cm]{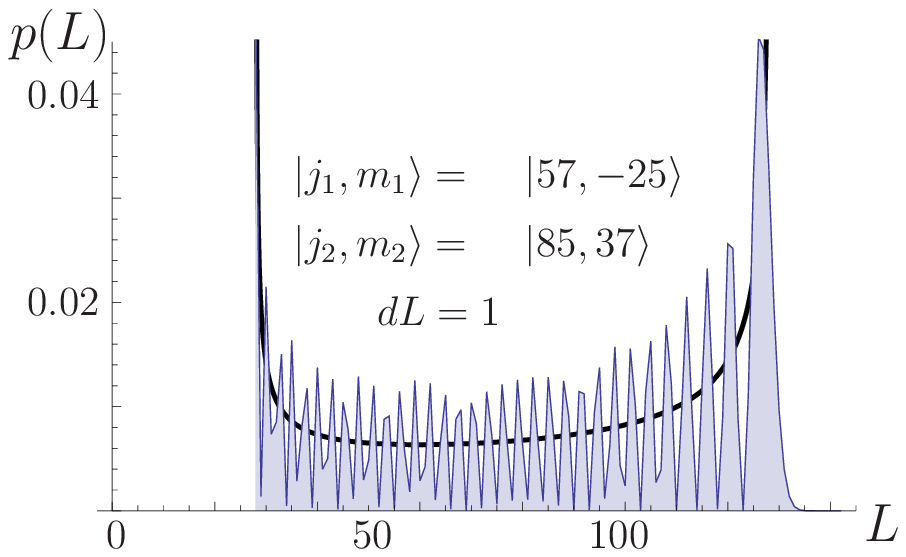}
    \includegraphics[width=7cm]{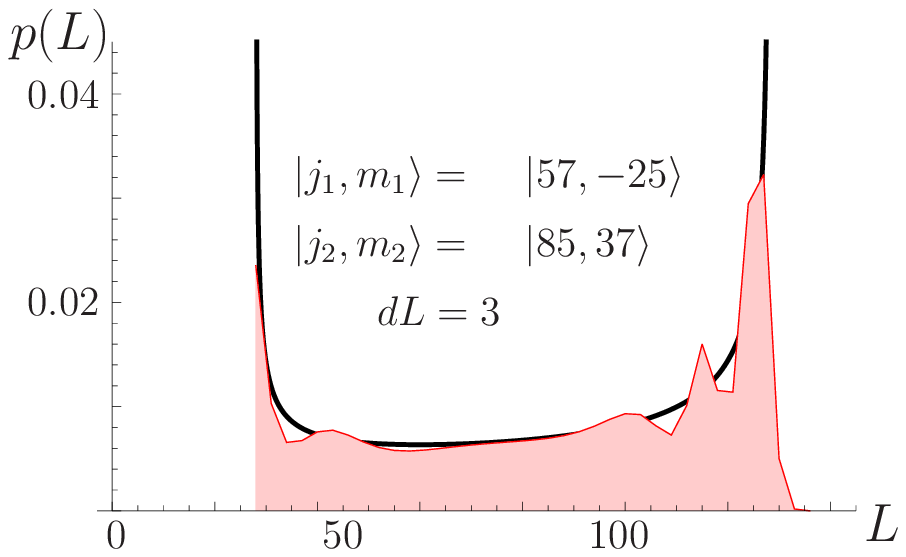}
    \caption{The role of resolution of the measuring device in allowing/preventing detection of accurate values of the Clebsch-Gordan coefficients (values of quantum numbers are given in the plot). If the~ resolution $dL$ (in units of $\hbar$) is~ such that $dL \lesssim 1$, then   
    the classical model does not reproduce the quantum values (left plot). In case $dL\gtrsim 1$ (right plot), the quantum values, averaged over $dL$, are well reproduced by the classical model. }
    \label{fig:fig4}
\end{figure}

Numerical values of the quantum mechanical Clebsch-Gordan coefficients are easily available. We used Mathematica to get their values and to plot their absolute values squared, as it is shown in Figs. \ref{fig:fig3}-\ref{fig:fig4}.

We will now compare the classical distribution of the total angular momentum with squares of the Clebsch-Gordan coefficients. We should stress that the classical angular momentum is a continuous quantity, whereas the quantum angular momentum is discrete. We have to, therefore, introduce a finite interval~ $dL$ of the continuous variable\ $L$. The probability density $p(L)$ multiplied by \  $dL$ should be compared with the quantum probability~ $p(J)$. A possible choice in our case and the easiest one, is to take $dL= \hbar$ -- the Planck constant. This allows to make a direct comparison of squares of the  Clebsch--Gordan coefficients and classical probability distribution of the total angular momentum.

It is seen from the plots that the classical probability distribution reproduces the general character of the quantum Clebsch-Gordan coefficients. This takes place even for relatively small values of angular momenta. Of course, the classical approach gives probabilities, as opposed to the quantum version, which gives probability amplitudes, and hence relative phases of the coefficients in addition to their absolute values.  

However, there are examples (shown by the black solid line in Fig. \ref{fig:fig3}), where the classical values do not represent properly the quantum values. This is due to their oscillations with a large amplitude and frequency equal to one unit of angular momentum $\hbar$. This does not mean that the classical approximation fails in these cases. 
This is one more illustration of the fact that the classical approximation gives average values only. If measurements of the angular momentum are precise enough to distinguish between values of $L$ that differ by $\hbar$, then differences between classical and quantum values can be found. If, however, the resolution is insufficient to detect values that differ by $\hbar$ then the classical picture suffices to describe the system.

This effect is illustrated in Fig. \ref{fig:fig4}, where two angular moment,\ $j_1=57$, $m_1=-25$\ and\ $j_2=85$, $m_2=37$\ are added. The~ dependence of the Clebsch-Gordan coefficients squared as functions of the total angular momentum are shown. The solid line represents the classical distribution given by \ref{classical_distribution}. If~ the resolution of measurement of the total angular momentum~ is~ 1 (meaning one quantum unit, hence $\hbar$), then the classical distribution differs from the quantum one. If, however, the resolution is $dL=3$, as shown in the right panel, the averaged quantum results are very close to the classical ones. Further discussion of this matter is given in Supplementary material (\ref{supcb}).

\section{Conclusions}

We have discussed classical interpretation of the Clebsch-Gordan coefficients, valid for large values of all three angular momenta involved. We have shown that in some cases the classical model gives a good approximation to the exact quantum values. This behavior is in fact expected.

The unexpected, in turn, was an appearance of conditions that are needed for the classical approach and the classical vector addition model to be valid. If these conditions are not fulfilled the classical vector addition model can be insufficient to find approximate values of the Clebsch-Gordan coefficients. This is most pronounced when the two added angular momenta have similar values. In this case Clebsch-Gordan coefficients exhibit rapid oscillations in the total angular momentum. These oscillations cannot be explained by any kind of classical model. 

Our results reveal an interesting feature of the classical limit of quantum mechanics. Not only all relevant physical quantities should have large values (large as compared to their single quantum units, like $\hbar$ in the studied case of angular momentum), but measurements of these physical quantity have to be taken into account as well. Quantum physics has to be used to describe results of measurements with resolution better than $\hbar$ whereas classical physics describes only measurements that average over intervals of angular momentum that are larger than the quantum unit.

This paper has shown aspects of angular momentum and addition of angular momenta that can be approximately described in the framework of classical physics. The analysis presented here should provide better understanding of quantum angular momentum physics. Classical analogues cannot, of course, explain quantum effects. They can, however, illustrate some general features of the system and their relation to classical physics.

\bibliography{ang}
\newpage
\newpage
\clearpage 
\newpage
\clearpage

\renewcommand{\thefigure}{S\arabic{figure}}
\setcounter{figure}{0}
\renewcommand{\thetable}{S\arabic{table}}
\setcounter{table}{0}
\renewcommand{\thesection}{S\arabic{section}}
\setcounter{section}{0}
\renewcommand{\thesubsection}{\Alph{subsection}}
\renewcommand{\theequation}{S\arabic{equation}}
\setcounter{equation}{0}
\renewcommand{\thepage}{S\arabic{page}}
\setcounter{page}{1}
\renewcommand{\appendixname}{}
\setcounter{footnote}{0}


{\huge Supplementary material }\\
\vspace{0.5cm}

\section{WKB approximation}
\label{supwkb}
We will now discuss some properties of spherical harmonics and their WKB (semiclassical) approximation. Spherical harmonics  $Y_{j,m}(\theta,\varphi)$ have the form $Y_{j,m}(\theta,\varphi)~=~\Theta_{j,m}(\theta)\exp(im\varphi)$. Functions $\Theta(\theta)$ satisfy equation:
\begin{equation}
    \frac{d^2\Theta_{j,m}}{d\theta^2}+\frac{\cos\theta}{\sin\theta}\frac{d\Theta_{j,m}}{d\theta}+ \left( j(j+1)-\frac{m^2}{\sin^2\theta} \right)\Theta_{j,m}=0
\end{equation}
Substitution $\Theta_{j,m}(\theta)=\frac{1}{\sqrt{\sin\theta}} T_{j,m}(\theta)$ gives the following equation for $T_{j,m}(\theta)$:
\begin{equation}
    \label{solveWKB}
    \frac{d^2\,  T_{j,m}}{d\theta^2}+\left(\frac{\cos^2\theta}{4\sin^2\theta}+\frac{1}{2} -\frac{m^2}{\sin^2\theta}+j(j+1)\right) T_{j,m}=0
\end{equation}
The Eq. (\ref{solveWKB}) has a form of the Schr\"odinger equation. The second derivative over $\theta$ plays the role of kinetic energy. Terms proportional to $j^2$ and $m^2$ are large in the semi-classical limit, while other terms are small. We will, therefore, skip these small terms.  Then, the equation takes the form:
\begin{equation}
\frac{d^2 T_{j,m}}{d\theta ^2 }+ \left(-\frac{m^2}{\sin^2{\theta}}+j^2 \right)T_{j,m}=0 
\end{equation}
and it is well suited for the WKB approximation. 
Let us note that the "kinetic energy" is positive only if $\theta$ is in the range between "classical turning points", i.e. $-\frac{|m|}{j}\le \sin\theta \le \frac{|m|}{j}$. 

In order to apply the WKB approximation we look for the solution in the form:
\begin{equation}
    T_{j,m}=\exp\left(i\, S(\theta)\right)+c.c.
\end{equation}
Next we expand $S(\theta)$ into power series in\ $j^{-1}$\ and take
into account that $m$ is of the same order of magnitude as $j$. The leading term is proportional to $j$, the next one is $j$ independent. Therefore, we get:
\begin{eqnarray}
    T_{j,m}(\theta)&=&\frac{1}{\Big(\sin^2\theta-\frac{m^2}{j^2}\Big)^{1/4}}\times \nonumber \\
    && \times \cos\left[j \left(\int  \sin{\theta'} d\theta^{\prime}\,  \sqrt{1-\frac{m^2}{j^2\sin^2\theta^{\prime}}}\right)\, -\phi\right]. \nonumber \\
    \label{notnormsemiT} 
\end{eqnarray}
The integration should be taken from the smaller "classical turning point". This formula does not give the overall sign of the function, the normalization, nor the overall phase $\phi$, these have to be found independently. 
The same formula as Eq. (\ref{notnormsemiT}), but restricted to $m=0$ only, can be found in \cite{Landau}. 

More careful analysis of the WKB approximation is given in \cite{Haake}. This result is:
\begin{eqnarray}
Y_{j,m}(\theta,\phi)&\simeq&  (-1)^{j-m} \frac{\left[ \frac{1}{\sin^2{\theta}-\frac{m^2}{\bar{J}^2}
}\right]^{1/4}}{\pi} \,  \, 
\cos{\left(\bar{J}\bar{S}_0-\frac{\pi}{4}\right)}\ e^{i\, m\phi},\nonumber \\
\\
\bar{S}_0&=&S_0(0,m)=\frac{m}{\bar{J}} \arccos{\left(\frac{\frac{m}{\bar{J}} \cot{\theta}}{\sqrt{1-\frac{m^2}{\bar{J}^2}}} \right)} + \nonumber \\
&&+ \arccos{\left(-\frac{ \cos{\theta}}{\sqrt{1-\frac{m^2}{\bar{J}^2}}} \right)} ,
\label{generatingF}
\end{eqnarray}
where $\bar{J}=j+\frac{1}{2}$ due to limit of large $j$ limit (or $\bar{J}$ as well).
This result is used in our numerical calculations. Differences between semiclassical and exact values are hardly visible for large $j$ in the allowed region of $\theta$.

\section{Oscillations of the Clebsch--Gordan coefficients}
\label{supcb}
In this part we will provide an explanation of  rapid oscillations of the CG coefficients seen in Fig. \ref{fig:fig3}.
\begin{figure}[b]
    \centering
    \includegraphics[width=6cm]{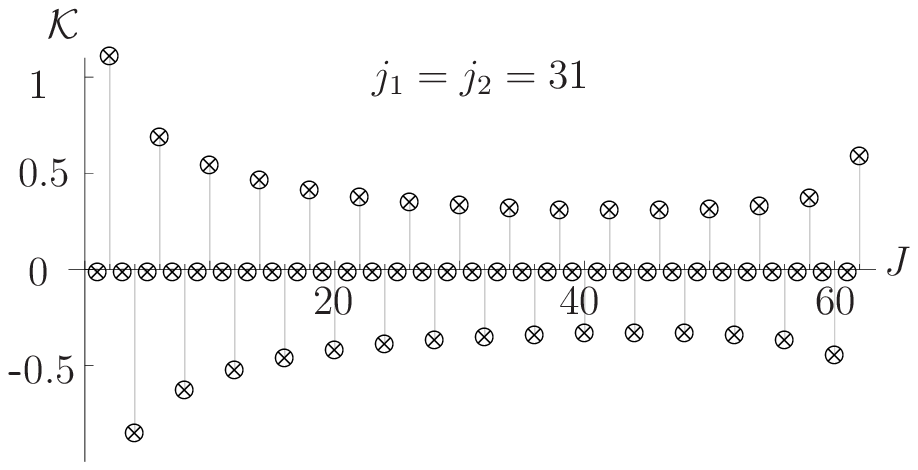}
    \includegraphics[width=6cm]{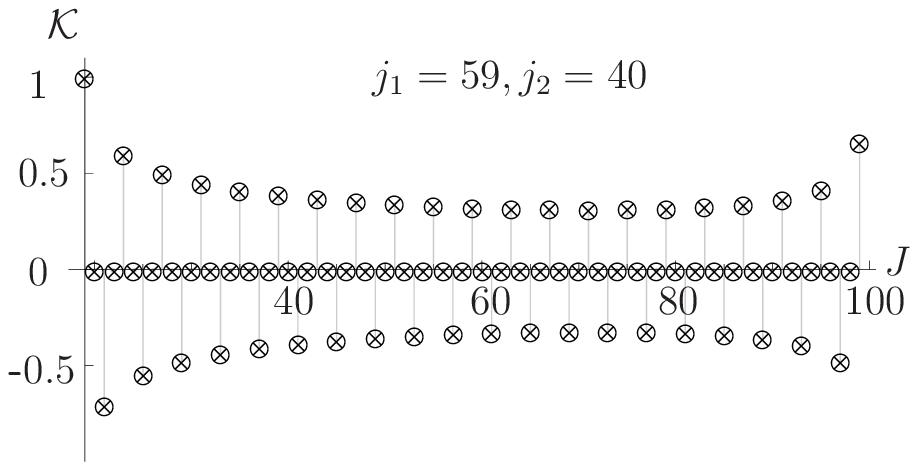}
    \caption{Coefficient $\mathcal{K}$ for various settings of angular momenta (values on the plot).}
    \label{fig:figS1}
\end{figure}
Classical approximation cannot account for this effect, semi-classical methods have to be used. In order to show the mechanism of oscillations we will now use the formula expressing Clebsch-Gordan coefficients in terms of integrals over spherical harmonics:
\begin{eqnarray}
  \int Y_{j_1,m_1}(\theta,\varphi)\, Y_{j_2,m_2}(\theta,\varphi)\, Y_{J,M}^{\ast}(\theta,\varphi)\ \sin{\theta}\,  d\theta\, d\varphi= && \nonumber \\
  \mathcal{K}\times \textrm{CB}(j_1,m_1,j_2,j_2;J,M), \qquad  \quad &&
    \label{three_harmonics}
\end{eqnarray}
where 
\begin{equation}
    \mathcal{K}=\sqrt{\frac{(2j_1+1)(2j_2+1)(2J+1)}{4\pi}}\ \Big(\begin{array}{ccc}
         j_1 & j_2 & J  \\
         0 & 0 & 0
    \end{array}\Big). 
\end{equation}
The Eq.(\ref{three_harmonics}) allows to determine the value of a CG coefficients only when the coefficient in front is not equal to zero. We~ are interested in the global features of the CG coefficients so this will not affect much the reasoning presented below.
Some of the values of the coefficient $\mathcal{K}$ are shown in Fig.\ref{fig:figS1}.\ It~ is clear that these coefficients do not depend strongly on $J$, however, they influence signs of CG.  
The first integration over $\varphi$ of the left hand side in the\ Eq.(\ref{three_harmonics}) 
gives a nonzero value only if\quad  $m_1+m_2=M$. The second integration over \ $\theta$ will have to be examined more closely.

\begin{figure}[h]
    \centering
    \includegraphics[width=5.2cm]{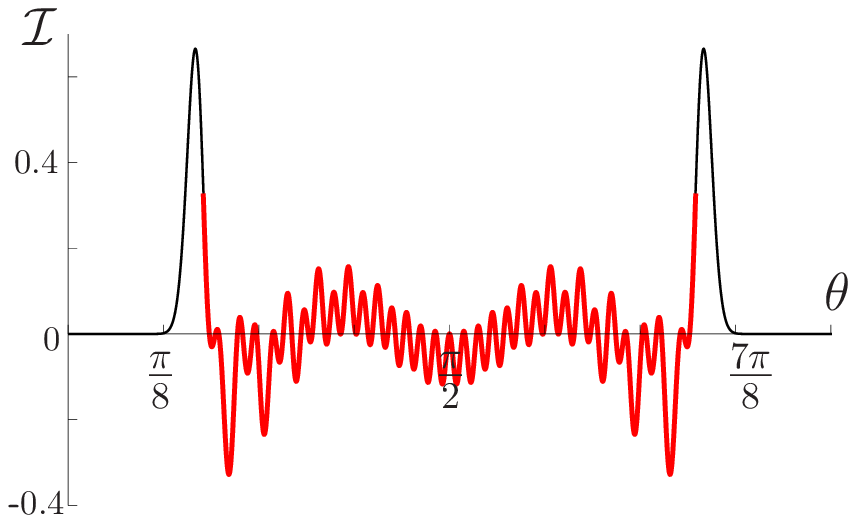}
    \includegraphics[width=5.2cm]{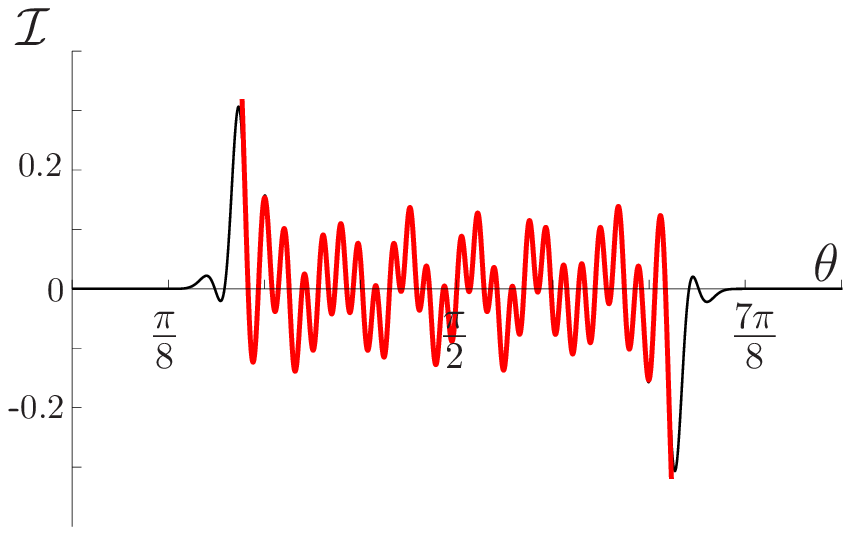}
    \includegraphics[width=5.2cm]{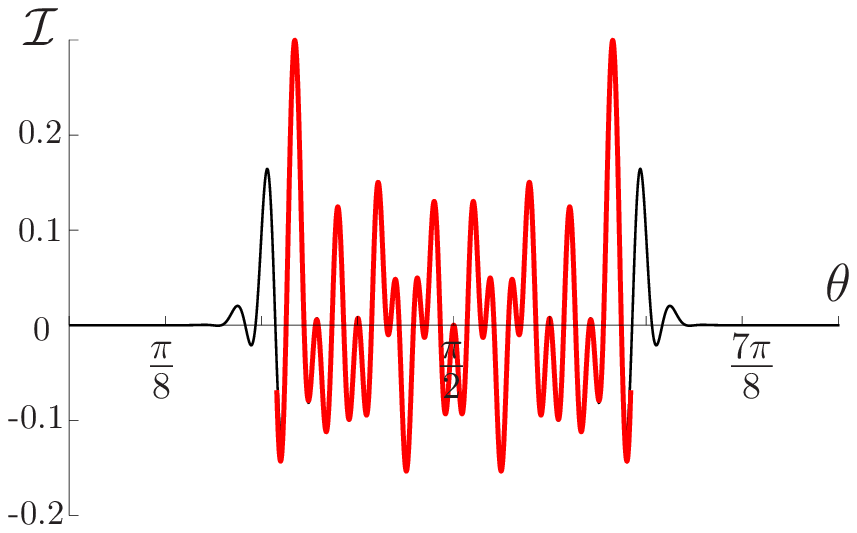}
    \caption{The integrand of Eq.(\ref{three_harmonics}), marked by $\mathcal{I}$, as a function of angle $\theta$ given for $j_1=j_2=31$ and $m_1=13$, $m_2=14$. The spherical harmonic results are shown by a black line, while their semi-classical counterpart results are shown in red. Parameters used for each plot, running from the left, are $J=56$, $J=43$ and  $J~=~36$. A~ perfect match between quantum and semiclassical functions is present for the whole $\theta$-region until turning points\  $\theta_0(M/J)\simeq \arcsin{(\frac{M}{J})}$ and\ $\theta_0(M/J)~=~ \pi- \arcsin{(\frac{M}{J})}$\   are reached. }
    \label{fig:figS2}
\end{figure}

 Fig. \ref{fig:figS2} gives the overall shape of the spherical harmonics, both exact and in the semi-classical approximation.  The largest mismatch to the exact values occurs in the vicinity of the classical turning points, i.e., \ for $\sin{\theta_0} \approx \sqrt{1-\frac{m^2}{j^2}}$. There, strictly speaking, the semi-classical approximation is not valid. What is seen, however, is a certain trend, the semiclassical function is large in this region.

Let us now discus a more complex case - the behavior of product of three semi-classical functions. With this, we will learn about the nature of the integrand,  in the same way as in case of Eq.(\ref{three_harmonics}).

 It can happen that all three functions in Eq.(\ref{three_harmonics}) have their turning points at about the same value of $\theta$. In this case, the~ value of the integral is determined by values of the integrand near the common turning points. One can also expect that the unusually large value of the integral occurs if all three functions are even (with respect to $\pi/2$). If one of the functions is odd and two other are even, than the values of the integral is unusually small -  the contribution from one turning point is almost exactly cancelled by the contribution from the other turning point.

\begin{figure}[t]
    \centering
    \includegraphics[width=8cm]{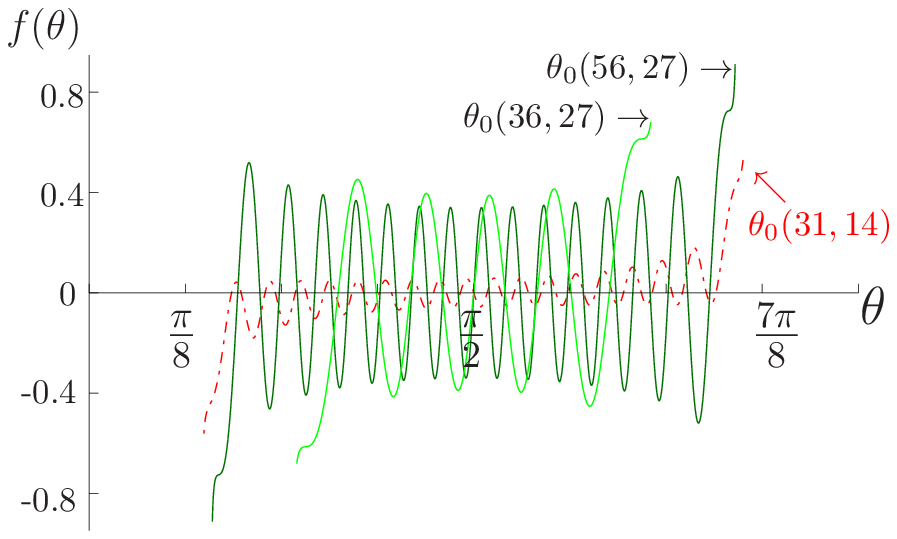}
    \caption{Functions contributing to integrand in the Eq. (\ref{three_harmonics}), shown in the range of angles between their own turning points. The red, dashed line indicates the product of harmonics in the semi-classical approximation $f(\theta)~=~Y_{j_1,m_1}(\theta,0)~\times~  Y_{j_2,m_2}(\theta,0)$\  for the values $j_1=j_2=31$ and $m_1=13$, $m_2=14$. In turn, both green lines show the behaviour of  the $f(\theta)=Y_{J,M}(\theta,0)$ function, which depends on the total angular momentum value $J$ (values on the plot). }
    \label{fig:figS3}
\end{figure}

\begin{figure}[b]
    \centering
    \includegraphics[width=6cm]{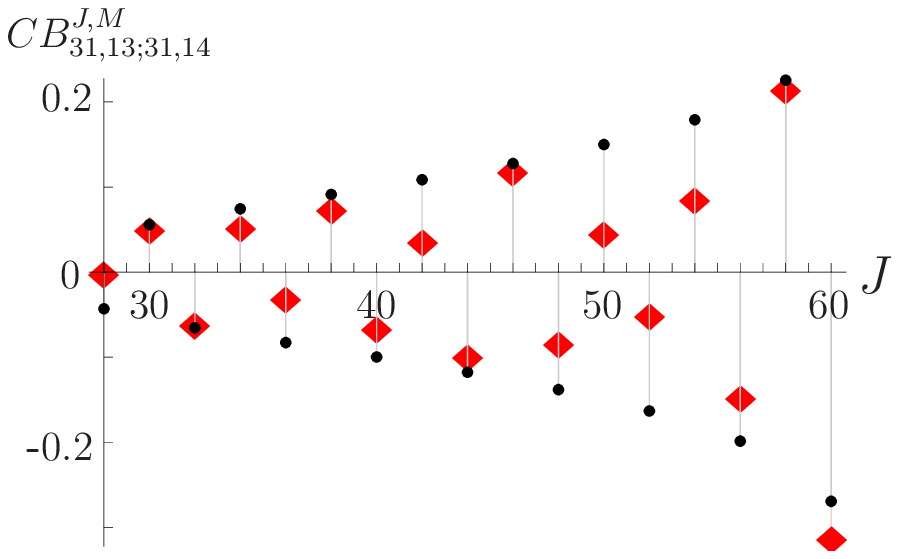}
    \caption{Clebsch-Gordan coefficients predictions obtained through integration of the left hand side of Eq.(\ref{three_harmonics}) and dividing by factor~ $\mathcal{K}$. The black dots correspond to pure quantum results, while the red diamonds are semiclassical one. This example concerns $j_1=j_2=31$ and $m_1=13$, $m_2=14$.}
    \label{fig:figS4}
\end{figure}

For better understanding, it is worth to have a look at the example in  Fig. \ref{fig:figS3}, where separate elements of the integrand are visualized. The red dashed line shows the behavior of the $Y_{j_1,m_1}(\theta,0)~\times~  Y_{j_2,m_2}(\theta,0)$ in the allowed range of\  $\theta$. Two~ green lines are semiclassical spherical harmonics $Y_{J,M}(\theta,0)$ for two different values of $J$. Results of integration of the product of the red and one of the green functions are proportional to CB$^{J,M}_{j_1,m_2;j_2,m_2}$ coefficients in semiclassical approach, however, their semiclassical values dependent strongly on the position of the turning points $\theta_0(M/J)$. This will be noted later, in Fig.\ref{fig:figS4}.  
 
The spherical harmonics, exact and in the semiclassical approximation, and also the $\mathcal{K}$ factor, change sign when $J$~ changes by $1$.  This is probably the reason why the integral in\ Eq.(\ref{three_harmonics}) changes its value when $J$ is changed by $1$,  and~ explains  the origin of the rapid oscillations of the~ CG coefficients.
 We~ should point out once more that this effect cannot be explained in the framework of the classical approximation.

 Having all this information we can understand the key result of this section, namely behavior of the Clebsch-Gordan coefficients Fig. \ref{fig:figS4}. It shows the computed values of CB coefficients, exact and in the semicalssical approximation for quantum numbers\ $j_1=j_2=31$ and\ $m_1=13$, $m_2=14$. Oscillations of the coefficients are clearly seen.

  Now, once again, we can state that the value and sign of the obtained results depend on the symmetry properties of the~ product of three spherical harmonics and on  relative positions of their classical turning points.
 
\clearpage
\end{document}